\pdfoutput=1
\documentclass[9pt,twocolumn,twoside,amsmath,amssymb]{osajnl}
\usepackage{cuted}  

\journal{ol} 

\setboolean{shortarticle}{true}

\title{Edge Modes of Scattering Chains with Aperiodic Order}
\author[1]{Ren Wang}
\author[2]{Malte Röntgen}
\author[2]{Christian V. Morfonios}
\author[3]{Felipe A. Pinheiro}
\author[2]{Peter Schmelcher}
\author[1,4,5,*]{Luca Dal Negro}

\affil[1]{Department of Electrical and Computer Engineering and Photonics Center, Boston University, Boston, MA 02215, USA}
\affil[2]{Zentrum für optische Quantentechnologien, Universität Hamburg, Luruper Chaussee 149, 22761 Hamburg, Germany}
\affil[3]{Instituto de Física, Universidade Federal do Rio de Janeiro, Rio de Janeiro-RJ 21941-972, Brazil}
\affil[4]{Department of Material Science and Engineering, Boston University, Boston, MA 02215, USA}
\affil[5]{Department of Physics, Boston University, Boston, MA 02215, USA}

\affil[*]{Corresponding author: dalnegro@bu.edu}

\ociscodes{(290.0290) Scattering; (290.4210) Multiple scattering; (290.5825) Scattering theory.}

\begin{abstract}
We study the scattering resonances of one-dimensional deterministic aperiodic chains of electric dipoles using the vectorial Green's matrix method, which accounts for both short- and long-range electromagnetic interactions in open scattering systems. We discover the existence of edge-localized scattering states within fractal energy gaps with characteristic topological band structures. Notably, we report and characterize edge-localized modes in the classical wave analogues of the Su-Schrieffer-Heeger (SHH) dimer model, quasiperiodic Harper and Fibonacci crystals, as well as in more complex Thue-Morse aperiodic systems. Our study demonstrates that topological edge-modes with characteristic power-law envelope appear in open aperiodic systems and coexist with traditional exponentially localized ones. Our results extend the concept of topological states to the scattering resonances of complex open systems with aperiodic order, thus providing an important step towards the predictive design of topological optical metamaterials and devices beyond tight-binding models.

\end{abstract}

\setboolean{displaycopyright}{true}

\begin{document}

\maketitle

\section{Introduction}
Analogues of topological insulating phases have been discovered in periodic photonic structures and quasiperiodic systems with modulated short-range coupling \cite{Rechtsman2013Nat,Noh2017NP,Levy2017EPJST,Verbin2015PRB,Bandres2016_PRX_6_011016_TopologicalPhotonic,Xiao2017_PRB_96_100202_PhotonicChern} in close correspondence with the transport of quantum waves in electronic materials. However, topological effects in dissipative (non-Hermitian) electromagnetic scattering systems that support long-range collective resonances have not been investigated.

In this paper, by systematically studying the spectral properties of the vectorial Green's matrix for linear chains of point dipoles with aperiodic order, we discover and characterize topological band structures and edge-localized resonances in open scattering systems.
In particular, our work shows that edge modes with topological dispersion appear in the complex spectrum of the scattering resonances as described by the vectorial Green's matrix, beyond the tight-binding approximation. Remarkably, we found that distinctive topological structures also appear in more complex deterministic aperiodic systems with Thue-Morse modulated coupling. We focus specifically on the optical analogues of the Su-Schrieffer-Heeger (SHH) and the Harper models, and extend our analysis to Fibonacci and Thue-Morse chains, which are primary examples of periodic, quasiperiodic, and deterministic aperiodic systems, respectively. Our results extend the concept of topological states to open scattering systems and provide an efficient tool for the predictive design of novel topological effects in aperiodic photonic structures.
\section{The Green's Matrix Method}
The Green's matrix method relies on the analysis of the spectra of the vectorial Green's matrix for three-dimensional vector scattering systems and provides invaluable insights into their general  physical properties. In fact, the vectorial Green's matrix also coincides with the kernel of the Foldy-Lax multiple scattering equations used to model arbitrary systems of coupled dipoles \cite{Guerin2006JOSAA,Christofi2016OL}. The method has been extensively used to understand wave transport in multiply scattering open random media \cite{Goetschy2011PRE,Rusek2000PRA,vanTiggelen1996PRE,Lagendijk1996PR} especially in conjunction with  Random Matrix Theory (RMT) \cite{Mehta2004}. Our group has recently applied this approach to understand the scattering properties of periodic, quasiperiodic, and deterministic aperiodic arrays of small nanoparticles \cite{dalNegro2016Crystal,Christofi2016OL}.
The elements of the $3N\times3N$ matrix are obtained from the relative positions of $N$ scattering dipoles \cite{Lagendijk1996PR}:
\begin{strip}
\begin{equation}
\begin{aligned}
\centering
\label{green}
\mathbf{G}_{nm}(k,\mathbf{r}_{nm})=\left\{
\begin{array}{l}
\frac{\exp (ikr_{_{nm}})}{i k r_{_{nm}}}
\text{\ }\left\{ \left[ {\mathbf U}-
\widehat{{\mathbf r}}_{_{nm}}
\widehat{{\mathbf r}}_{_{nm}}\right] -\left(
\frac{1}{ikr_{_{nm}}}+\frac{1}{(kr_{_{nm}})^{2}}\right)
\left[ {\mathbf U}-3
\widehat{{\mathbf r}}_{_{nm}}\widehat{{\mathbf r}}_{_{nm}}\right]
\right\}
\text{\ \ \ for \ \  }n\neq m,\\
0\ \ \ \ \ \ \ \ \ \ \ \ \ \ \ \ \ \ \ \ \ \ \ \ \ \ \
 \ \ \ \ \ \ \ \ \ \ \ \ \ \ \ \ \ \ \ \ \ \ \ \ \ \ \
\ \ \ \ \ \ \ \ \ \ \ \ \ \ \ \ \ \ \ \ \ \ \ \ \ \ \ \
\ \ \ \ \ \ \ \ \ \ \ \ \ \ \ \ \ \ \ \ \ \
\text{for \ \ }n=m,
\end{array}
\right.
\end{aligned}
\end{equation}
\end{strip}
where $\mathbf{G}_{nm}$ is the $3\times3$ block element of the vectorial Green's matrix, the integers $n, m \in {1, 2, 3, ..., N}$ label particles's positions, $k$ is the optical wavenumber, $\mathbf{U}$ is the $3\times3$ identity matrix, $\widehat{{\mathbf r}}_{_{nm}}$ is the unit vector position from the $n$-th to $m$-th particle, and its magnitude is $r$. The Green's matrix describes the coupling of each scatterer to all the other scatterers in an arbitrary system through the propagation of electromagnetic vector waves. The vectorial Green's matrix in (\ref{green}) is the sum of three components, each describing electromagnetic interactions proportional to $1/r_{nm}$, $1/r^2_{nm}$, and $1/r^3_{nm}$ corresponding to long-range, intermediate range, and short-range coupling, respectively.
Since a Green's matrix is non-Hermitian, its eigenvalues are complex and they entirely characterize the scattering resonances of the system. Specifically, the real and the imaginary parts of the complex eigenvalue $\Lambda_{\alpha}$ are related to the relative decay rate $(\Gamma_{\alpha}-\Gamma_{0})/\Gamma_{0}$ and relative energy position $(\omega_{\alpha}-\omega_{0})/\Gamma_{0}$ of a scattering resonance, respectively \cite{Rusek2000PRA,dalNegro2016Crystal,Christofi2016OL} ($\alpha \in {1, 2, 3, ..., 3N}$). In the following sections, we will focus on the distribution of energy states of the resonances of chains of dipole scatterers with different positional order, and retrieve topological information directly from the eigenvalues of the Green's matrix. Although we present results on chains with optical density $\rho=10$ (10 dipoles per wavelength), our main conclusions remain valid even in lower density chains down to approximately $\rho=1$.
\begin{figure}
\centering
 \includegraphics[scale=0.28]{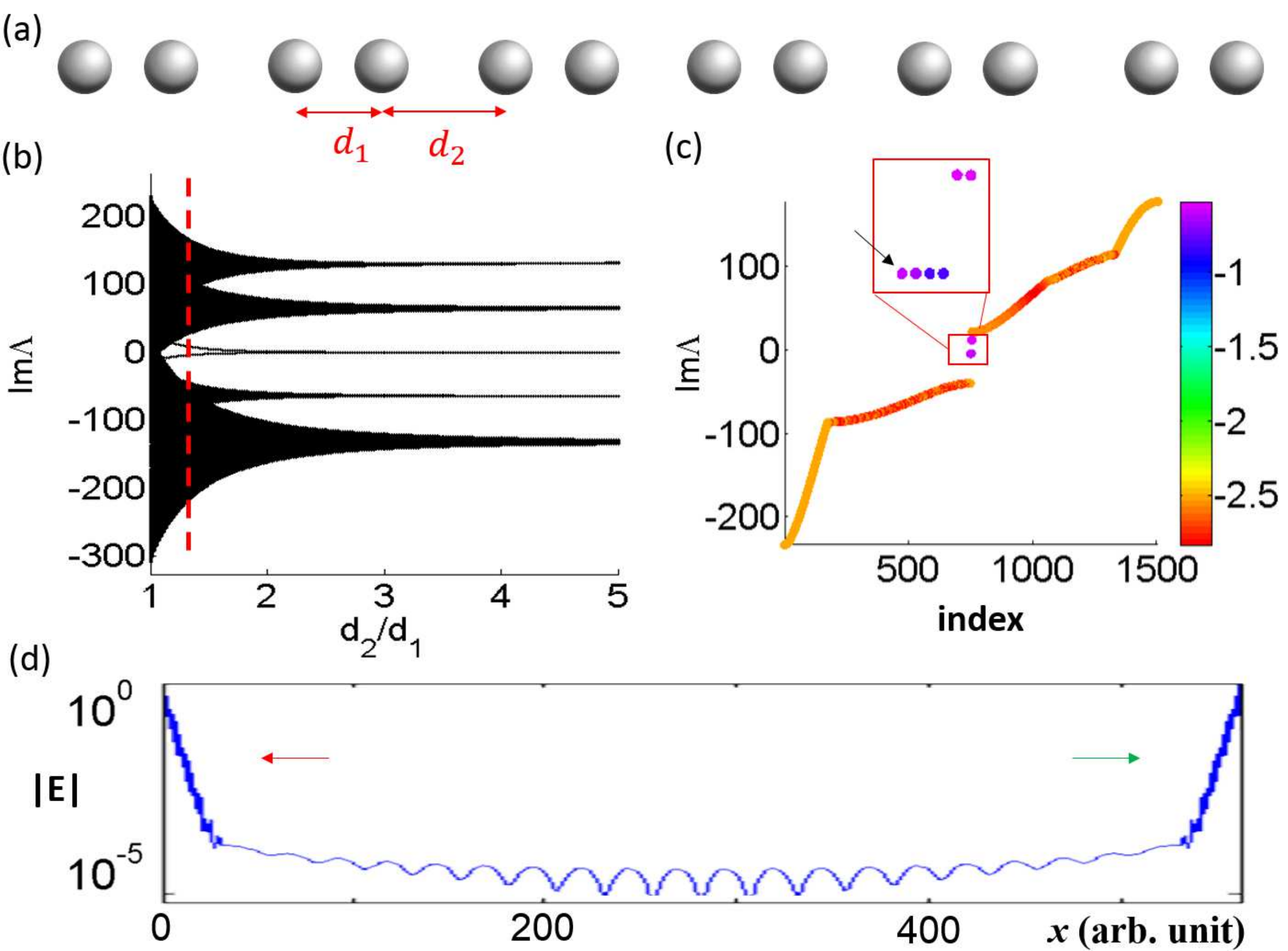}
 \caption{\label{fig1} (a) Dimer chain of identical dipole scatterers with alternating distance modulation. (b) Topological band-structure from the vectorial Green’s matrix of a 500-dipole chain. (c) Energy spectrum for $d_{2}/d_{1} = 1.25$ (red line in (b)), and the inset shows the 6 edge modes inside the band gap. Color-bar shows $log_{10}(IPR)$. (d) A representative edge mode in the gap.}
\end{figure}
\section{Results and Discussion}
First we consider a doubly periodic dimer chain, which is the photonic counterpart of the Su-Schrieffer-Heeger (SHH) model of a topological insulator \cite{Asboth2016}. This model describes spinless fermions hopping on a one-dimensional lattice with staggered hopping amplitudes. In our photonic implementation we identify the SHH hopping amplitudes with two separate values of inter-particle separations, $d_{1}$ and $d_{2}$, which alternate along the chain as illustrated schematically in Fig. \ref{fig1}(a). The periodic variation in the inter-dipole separations results in a modulation of the electromagnetic interaction that gives rise to distinctive gaps and topological effects. However, differently from the standard tight-binding description, in our photonic implementation the vector dipoles interact not only through short-range coupling, but also via the intermediate- and long-range contributions of the Green's matrix. In Fig. \ref{fig1}(b), the imaginary part of the eigenvalues, which is proportional to the energy position of the scattering resonances, is plotted for the dimer chain against the ratio of the two inter-particle distances, $d_{1}/d_{2}$, which ranges from 1 to 5. In close analogy with the case of the electronic SHH dimer case \cite{Asboth2016}, this quantity plays the role of an effective topological parameter. As it is varied, there appear in Fig. \ref{fig1}(b) energy gaps that are traversed by band-gap states near $Im\Lambda=0$. We will now show that these band-gap states correspond to edge-localized modes.
In particular, we focus for concreteness on the spectrum of the dimer structure with $d_{1}/d_{2} = 1.25$, which is highlighted by the vertical red line in Fig. \ref{fig1}(b), and plot in Fig. \ref{fig1}(c) the energies of its scattering resonances. We observe  one clear band-gap around $Im\Lambda=0$ containing 6 states.
\begin{figure}
\centering
 \includegraphics[scale=0.28]{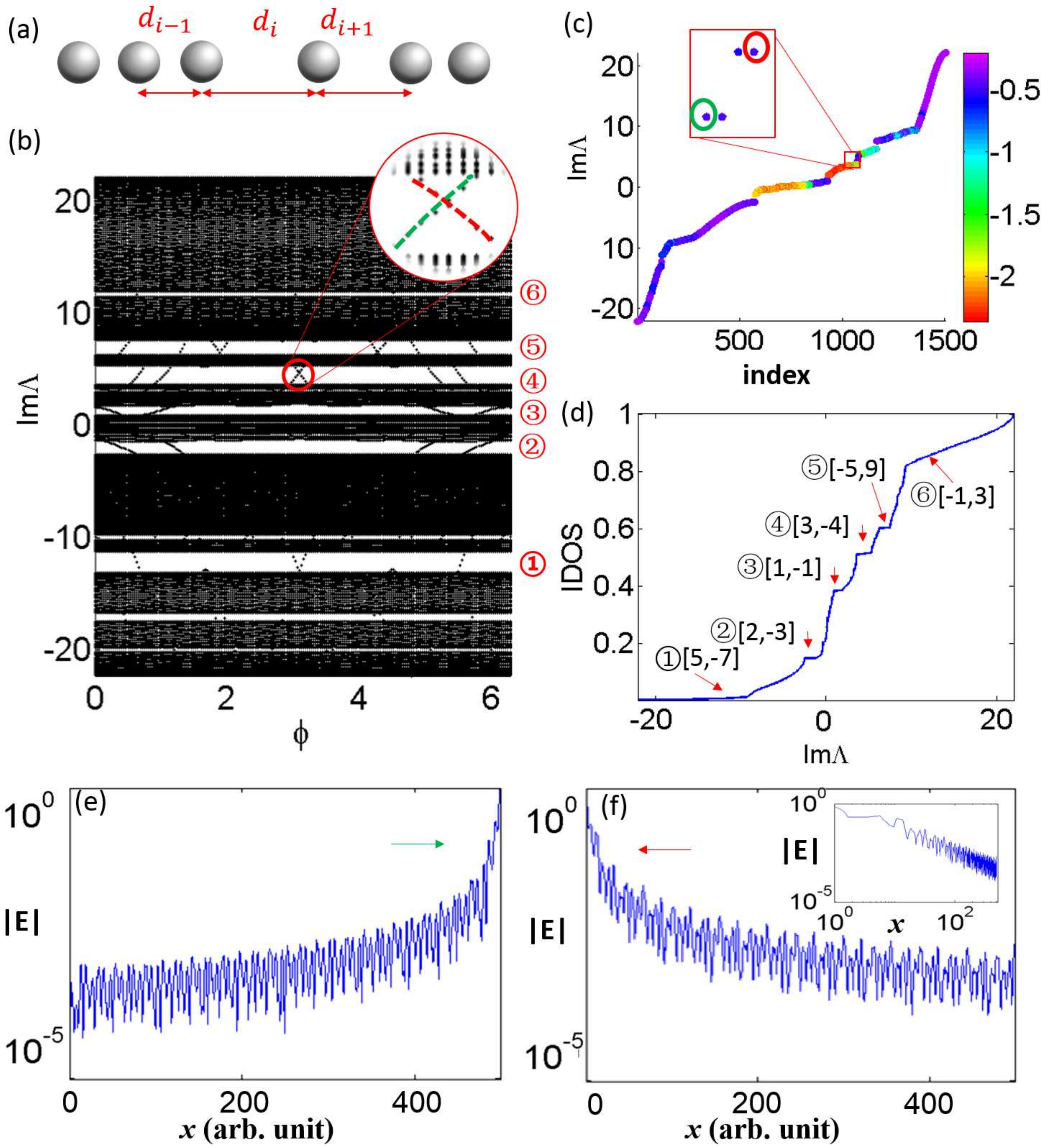}
 \caption{\label{fig2} (a) Harper chain with distance modulation; (b) Topological band-structure obtained with with the vectorial Green's matrix for the Harper chain with 500 particles. The inset is a zoom-in around the crossing of the edge states. (c) Energy levels for $\phi = 3.05$, and the edge-state in the gap. (d) Corresponding IDOS and labeling of gaps  by the two integers [p,q] as in Eq. \ref{gaplabeling}. (e) and (f) are two edge modes in the gap. The inset of (f) shows the log-log plot of the field profile.}
\end{figure}
The inset of Fig. 1(c) magnifies the energy region around the band-gap states, which separate into 2 distinct energies. These band-gap states originate from the 3 polarizations of the vectorial Green's matrix and are doubly degenerate. In order to better understand their spatial localization properties, we compute the inverse participation ratio (IPR) for the eigenmodes defined as \cite{Goetschy2011PRE}:
\begin{equation}\label{IPR}
IPR(\Lambda_{\alpha})=\frac{\sum_{\beta=1}^{3N}|\textbf{e}_{\beta}(\omega_{\alpha})|^{4}}{[\sum_{\beta=1}^{3N}|\textbf{e}_{\beta}(\omega_{\alpha})|^{2}]^{2}} \ ,
\end{equation}
\noindent
where $\textbf{e}_{\beta}(\omega_{\alpha})$ is the (normalized) eigenmode component of the Green's matrix corresponding to the mode of energy $\omega_{\alpha}$. The eigenmodes are normalized so that the obey the condition $\sum_{\beta}|\textbf{e}_{\beta}(\omega_{\alpha})|^{2}=1$.
The IPR measures the degree of spatial extent of each eigenmode in the system. For example, an eigenmode that extends over all the $N$ scattering centers is characterized by a low value of $IPR\simeq{1/N}$, while an eigenmode localized at a single dipole of the chain has a large $IPR=1$. Fig. \ref{fig1}(c) is color-coded to reflect the value of the IPR for each eigenstate in a logarithmic scale, and demonstrates that the modes residing within the gap region display the largest IPR values.
We plot in Fig. \ref{fig1}(d) the spatial distribution of one representative edge-state in the selected gap region, which displays a large amplitude localized at both edges of the chain. The remaining  band-gap states have almost identical edge-localized profiles.
\begin{figure}
\centering
 \includegraphics[scale=0.28]{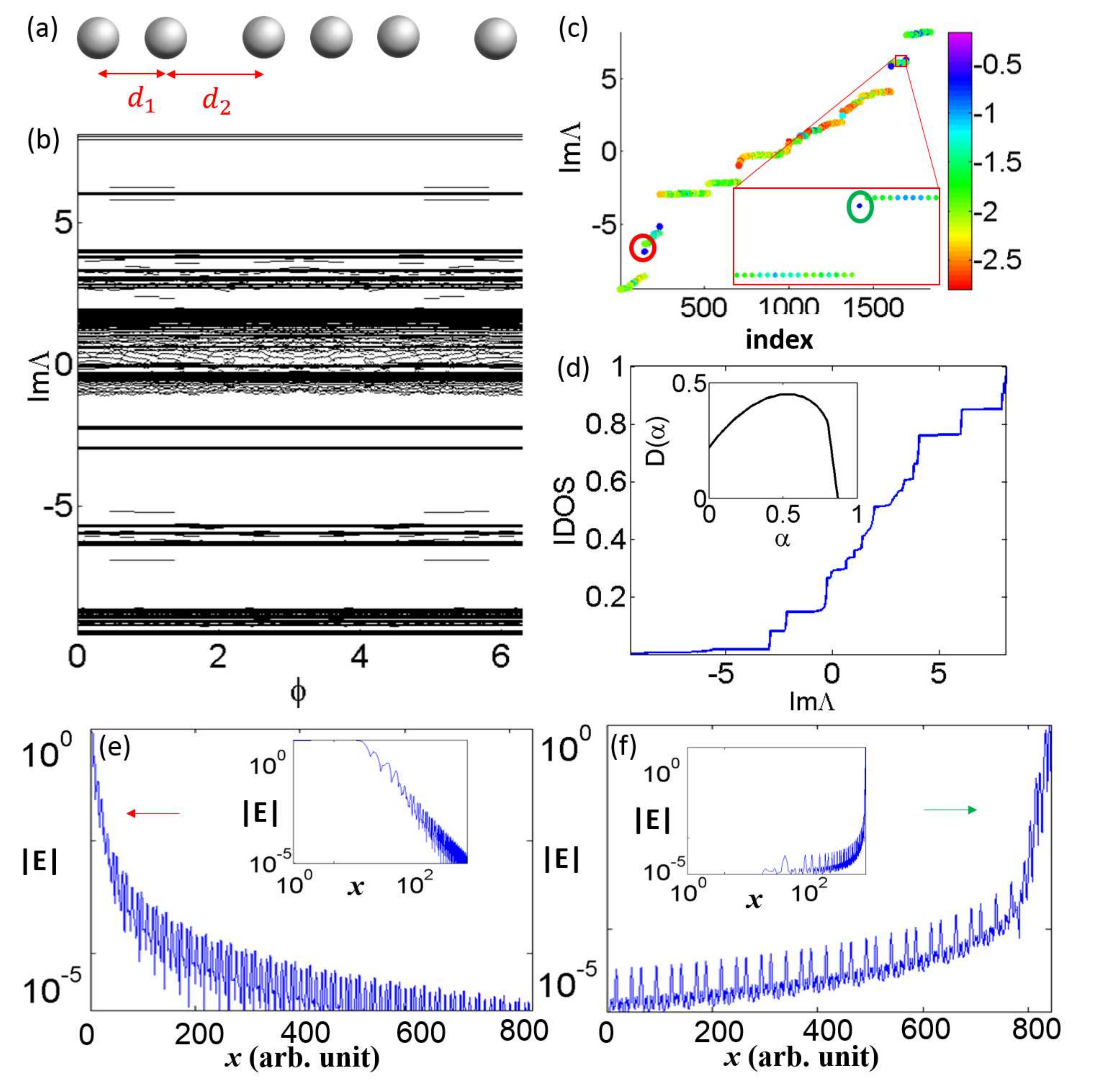}
\caption{\label{fig3} (a) Schematic illustration of a Fibonacci chain of point scatterers with binary distance modulation $d_{2}/d_{1} = 2$; (b) Topological band-structure for the energy levels for a 610-particle Fibonacci chain.  (c) Energy spectrum for the $\phi = 1$ case showing an edge-state in the gap. (d) Corresponding IDOS for the $\phi = 1$ case. The inset plots the singularity spectrum D($\alpha$) of the gap width distribution. (e) Edge mode with power-law decay and (f) edge mode with exponential localization (insets are in double-logarithmic scale).}
\end{figure}

We now apply the same methodology to more complex structures with aperiodic order. Tight-binding models for deterministic aperiodic potentials based on the Harper Hamiltonian or the Fibonacci quasi-periodic potential have attracted a considerable interest in condensed matter theory \cite{Macia2014ISRN,Hofstadter1976PRB,Han1994PRB} and have also been recently studied in optics \cite{Verbin2015PRB,Lucic2015JOSAB}. Here we address these systems using the vectorial Green's matrix method for the first time.
We consider an optical scattering chain defined by the inter-particle separations:
\begin{equation}\label{harper}
d_{n} = d_{0}\left[1-\varepsilon\cos\left(\frac{2\pi}{n\tau} + \phi\right)\right],
\end{equation}
\noindent
where $d_{n}$ is interpreted as the $n$-th inter-particle distance, $d_{0} = 1$ is a scaling constant, $\varepsilon = 0.5$ is the coefficient that controls the strength of the modulation, $\tau = (\sqrt{5}+1)/2$ is the golden ratio, and $\phi$ is the topological parameter. Since $\tau$ is irrational, the spatial modulation is quasiperiodic.
Moreover, when $\phi$ varies continuously between $0$ and $2\pi$, Eq. (\ref{harper}) describes an adiabatic deformation of the chain. The modulation defined in (\ref{harper}) is isomorphic to the Aubry-Andr\'{e} or Harper model describing lattice systems in the presence of a gauge field \cite{Hofstadter1976PRB,Macia2014ISRN}.
A schematic illustration of a portion of the investigated Harper chain is shown in Fig. \ref{fig2}(a). In Fig. \ref{fig2}(b), we display the computed topological bandstructure obtained by  varying the parameter $\phi$ from 0 to $2\pi$. The bandstructure features a very rich distribution of energy band-gaps in the scattering spectra that are crossed by different numbers of edge-localized modes. In addition, at $Im\Lambda\simeq 5$ and when $\phi\simeq\pi$ we observe a crossing between two band-gap states that gives rise to a crossing, similar to a Dirac point (see magnified picture in the inset). The two crossing states are strongly localized at the two opposite edges of the chain.
In Figs. \ref{fig2}(c)-(f) we study the energy spectrum for a Harper chain with $\phi =3.05$, which is near the crossing point of two edge states. In Fig. \ref{fig2}(c), we plot the IPR color-coded imaginary part of the scattering eigenvalues against their indices in the chain, and the inset shows four strongly localized edge-states (doubly degenerate) inside the band-gap. Based on the imaginary part of the eigenvalues, we have also calculated the IDOS for the Harper chain, which is shown in Fig. \ref{fig2}(d). The IDOS displays a characteristic staircase structure where each plateau indicates the presence of an energy gap in the spectrum. We have identified each gap by applying the gap-labeling theorem using the relation \cite{Baboux2017PRB}:
\begin{equation}\label{gaplabeling}
N = p + q\tau,
\end{equation}
where $\tau$ is the golden number, $p$ and $q$ are integer labels. Besides, $q$ can be interpreted as the topological Chern numbers of the labelled gaps \cite{Baboux2017PRB}.
In Figs. \ref{fig2}(e) and \ref{fig2}(f) we show the spatial profiles of two representative band-gap states that are strongly localized at the two edges of the chain. However, unlike the dimer case, the edge-localized states of the Harper chain display a characteristic power-law amplitude localization (see log-log plot in the inset of \ref{fig2}(f)) attributed to the fractal nature of the underlying eigenmodes.

We now discuss the scattering resonances of one-dimensional Fibonacci chains, which have been widely investigated due to their unique light transport and localization properties \cite{dalNegro2003PRL}. Moreover, such canonical structures recently attracted significant attention in relation to the topological properties of short-range coupled photonic waveguide systems \cite{Levy2017EPJST,Baboux2017PRB}. A one-dimensional Fibonacci chain can be obtained as a special case of the Harper model equation \cite{Kraus2012_PRL_109_116404_TopologicalEquivalence,Baboux2017PRB}:
\begin{equation}\label{Fibonacci}
  d_{n} = sgn\left[\cos\left(\frac{2n\pi}{\tau} + \phi\right) - \cos\left(\frac{\pi}{\tau}\right)\right],
\end{equation}
where $sgn$ denotes the sign operator of the argument that outputs only the binary values $\pm1$, and $\tau$ is the golden mean. As a result, by varying the topological parameter $\phi$, equivalent representations of the Fibonacci chains can be generated.
In Fig. \ref{fig3}(a) we illustrate a portion of the Fibonacci binary sequence. In Fig. \ref{fig3}(b) we vary the topological parameter $\phi$ from $0$ to $2\pi$ and obtain the topological band diagram that features a rich distribution of energy band-gaps and gap-localized states. However, unlike in the case of the Harper chain, the structure of the Fibonacci chain flips abruptly when continuously varying the topological parameter $\phi$ due to the effect of the $sgn$ operation in Eq. \ref{Fibonacci}. These sudden structural transitions are known as phason flips \cite{Levy2016arXiv}.
\begin{figure}[h]
\centering
 \includegraphics[scale=0.30]{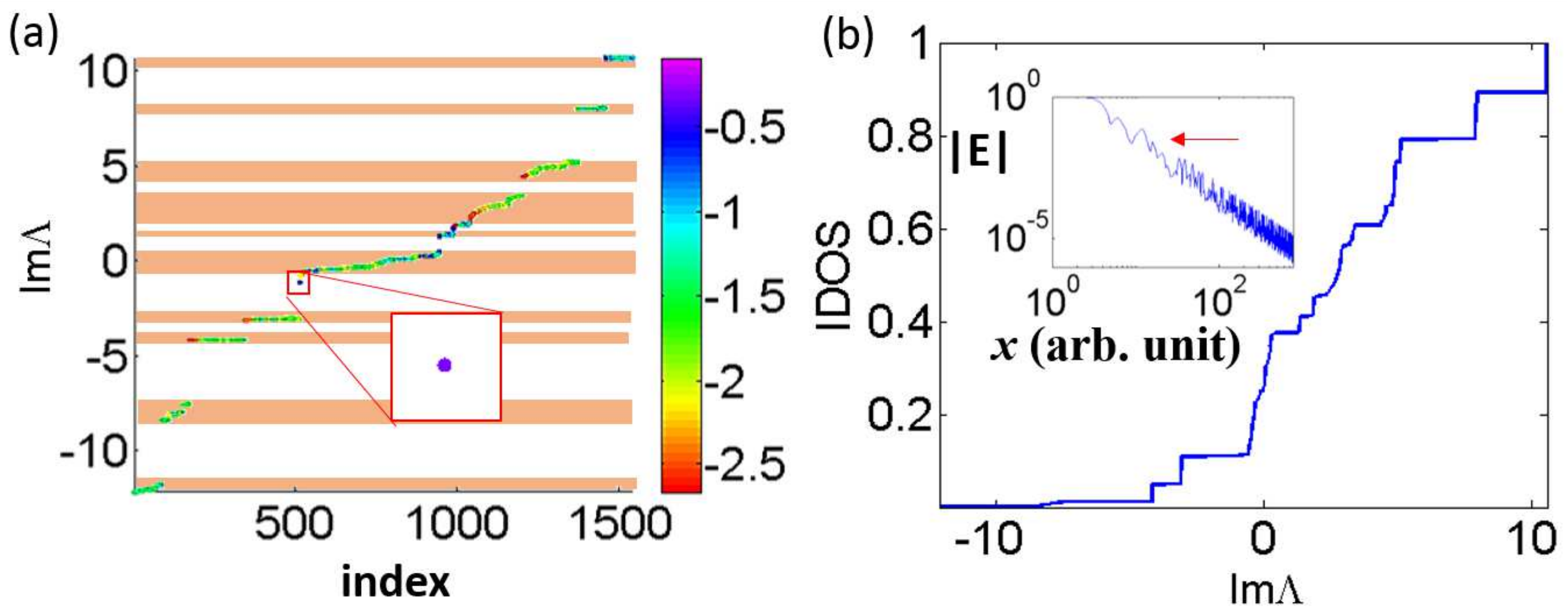}
 \caption{\label{fig4} (a) Green’s matrix energy spectrum and band-structure for a 512-particle Thue-Morse linear chain, with binary distance modulation $d_{2}/d_{1} = 2$. The inset shows a magnified view of a band-gap state. (b) IDOS of the Thue-Morse chain, and edge-localized eigenstate inside the band gap (inset).}
\end{figure}
\par By choosing the value $\phi=1$, we show in Fig. \ref{fig3}(c) the distribution of the imaginary parts of the eigenvalues, and identify two edge-localized states residing in the band-gaps circled (see inset). The corresponding IDOS is displayed in Fig. \ref{fig3}(d), and consists of a fractal staircase of band-gap widths. The fractal nature of the gap-width distribution is demonstrated by the singularity spectrum shown in the inset, obtained by using the Wavelet Transform Modulus Maxima (WTMM) method described in \cite{dalNegro2017SR}. In Fig. \ref{fig3}(e) we show the spatial distribution of the edge state highlighted by the red circle in Fig. \ref{fig3}(c), while in Fig. \ref{fig3}(f) we display the edge state inside the green circle in Fig. \ref{fig3}(c). It is evident that both states are strongly localized on the edges of the chain. However, we also notice that the amplitude in Fig. \ref{fig3}(e) decays as a power-law (see double logarithmic scale in the inset) while the state in Fig. \ref{fig3}(f) is exponentially localized, similarly to the dimer chain. Such a coexistence of two different localization regimes in a single physical system is a novel phenomenon driven by the long range correlation of the electromagnetic interactions, as captured by the vectorial Green's matrix.
\par The results obtained so far demonstrate that the presence of edge modes is a generic feature of electromagnetically coupled, one-dimensional chains with aperiodic order. However, topological parameters for more complex structures beyond quasi-periodic order, such as one-dimensional Thue-Morse chains, have not yet been reported. We address this challenge directly by computing the energy spectrum of the Thue-Morse chain, which unveils the presence of edge-localized band-gap states. This is achieved in Fig. \ref{fig4}(a) where the imaginary part of the Thue-Morse eigenvalues for a linear chain with $N = 512$ dipoles. We observe the presence of a single state in the band-gap near $Im\Lambda = 0$.
Here, we are not able to obtain a full topological bandstructure due to the lack of a known topological parameter for Thue-Morse systems. However, we report clear edge-localized states with power-law amplitude decay in the Thue-Morse energy gaps, similar to the Harper or Fibonacci chains. Finally, in Fig. \ref{fig4}(b) we show the IDOS of the Thue-Morse chain, which also displays a complex staircase structure. Further work beyond the scope of this contribution is needed in order to understand the nature of edge-localization in Thue-Morse structures.

\section{Conclusion}
In conclusion, through the study of the vectorial Green's matrix spectra of electromagnetic scattering systems including analogs of SHH and Harper models, as well as aperiodic Fibonacci and Thue-Morse chains, we have unambiguously identified and characterized scattering edge modes. In particular, we have found crossing of modes localized on opposite edges inside band gaps, analogously to the crossing of edge states with opposite chirality in topological insulators. We have also discovered a novel power-law edge-mode localization in deterministic aperiodic one-dimensional scattering systems. Notably, Fibonacci chains are found to support both power-law and exponentially decaying edge states. 
Our work extends the concept of topological states to open electromagnetic scattering systems and fully demonstrates the potential of the vectorial Green's matrix method for the engineering of novel topological phenomena in collectively coupled electromagnetic structures and aperiodic metamaterials.

\section{Funding} This work was supported by the Army Research Laboratory through the Collaborative Research Alliance (CRA) for MultiScale Multidisciplinary Modeling of Electronic Materials (MSME) under Cooperative Agreement Number W911NF-12-2-0023. F.A.P. thanks the Royal Society-Newton Advanced Fellowship (Grant No. NA150208), CNPq, and FAPERJ for financial support. M.R. gratefully acknowledges financial support by the 'Stiftung der deutschen Wirtschaft'.


\bibliography{EdgeStatePRB_references}

\bibliographyfullrefs{EdgeStatePRB_references}

\end{document}